\documentclass[letterpaper, 10 pt, conference]{ieeeconf}
\IEEEoverridecommandlockouts
\usepackage[utf8]{inputenc}
\usepackage{graphicx} 
\usepackage{lipsum, babel}
\usepackage{xcolor}
\usepackage{soul}

\usepackage{amsmath}
\usepackage{tikzscale}
\usepackage{pgfplots}
\usepackage{tikz}
\usepackage{amssymb}
\usepackage{hyperref}
\pgfplotsset{compat=1.18}

\usepackage{subcaption}

\makeatletter
\def\@endtheorem{\endtrivlist}
\makeatother

\newcommand{\calE}{{\mathcal E}}    
    
\newcommand{\calG}{{\mathcal G}}

\newcommand{\calN}{{\mathcal N}}

\newcommand{\calS}{{\mathcal S}}

\newcommand{\bbE}{{\mathbb E}}

\newcommand{\bbP}{{\mathbb P}}    
    
\newcommand{\bbR}{{\mathbb R}}

\newcommand{\subg}{{\mathcal{SG}}}
\newcommand{\subex}{{\mathcal{SE}}}

\newcommand{\Ee}{{\rm e}}

\newtheorem{theorem}{Theorem}
\newtheorem{corollary}{Corollary}
\newtheorem{lemma}{Lemma}
\newtheorem{proposition}{Proposition}

\newtheorem{assumption}{Assumption}
\newtheorem{definition}{Definition}

\newtheorem{remark}{Remark}

\AtBeginEnvironment{theorem}{\itshape}
\AtBeginEnvironment{corollary}{\itshape}
\AtBeginEnvironment{lemma}{\itshape}
\AtBeginEnvironment{proposition}{\itshape}
\AtBeginEnvironment{problem}{\itshape}
\AtBeginEnvironment{assumption}{\itshape}
\AtBeginEnvironment{definition}{\itshape}

\def\cov{{\rm cov}}

\title{Concentration-Based Probabilistic Reachability for Linear Systems under Multiplicative Uncertainty and Additive Noise}
\author{Kaouther Moussa$^{1,2,*}$, Kaito Ito$^{3,*}$, and Hideaki Ishii$^{3}$
\thanks{$^*$ K. Moussa and K. Ito contributed equally to this work.}
\thanks{This work was supported in part by JSPS KAKENHI Grant Numbers JP24K17297, JP23K22778, by JST ASPIRE Grant Number JPMJAP2402, and by the Clinical project, funded by the ANR (French National Research Agency) under grant ANR-24-CE45-4255.}
\thanks{$^1$ K. Moussa is with UPHF, CNRS, UMR 8201 - LAMIH, F-59313 Valenciennes, France. {\tt \small Kaouther.Moussa@uphf.fr}}%
\thanks{$^2$ K. Moussa is also with INSA Hauts-de-France, F-59313, Valenciennes, France.}%
\thanks{$^3$ K. Ito and H. Ishii are with the Department of Information Physics and Computing, The University of Tokyo, Tokyo, Japan. {\tt \small \{kaito\_ito,hideaki\_ishii\}@ipc.i.u-tokyo.ac.jp}}%
}

\begin{document}

\maketitle
\begin{abstract}
This paper investigates probabilistic reachability for discrete-time linear stochastic systems subject to bounded multiplicative uncertainty and potentially unbounded additive noise. To address the conservatism of reachable sets constructed solely from covariance information, we develop two complementary approaches based on sub-Gaussian and sub-exponential matrix moment-generating-function characterizations. The sub-Gaussian approach relies on a worst-case envelope over the uncertainty support, whereas the sub-exponential approach exploits expectation-based propagation of the matrix variance proxy. Both approaches are used to construct probabilistic reachable sets with explicit probability guarantees, considering ellipsoidal confidence regions and half-space bounds. For the latter, we derive a necessary and sufficient condition under which the sub-exponential chance-constraint margin is no larger than its sub-Gaussian counterpart. 
Numerical results show that the proposed bounds can be less conservative than those based on the covariance, and illustrate the complementary nature of the two approaches across probability levels.
\end{abstract}

\section{Introduction}
\label{sec:introduction}

Probabilistic reachable sets (PRS) provide a means to quantify state
uncertainty and enforce chance constraints in stochastic control.
Kofman \emph{et al.}~\cite{KofmanAUT12} studied probabilistic set
invariance and ultimate boundedness, including noise with
unbounded support. In stochastic model predictive control (SMPC),
Hewing \emph{et al.}~\cite{Hewing2020} used PRS for constraint
tightening and recursive feasibility under additive noise,
while Fiacchini and Alamo~\cite{Fiacchini2021} constructed
probabilistic reachable and invariant sets for systems affected by correlated noise.
More recently, Ao \emph{et al.}~\cite{Ao2025} exploited sub-Gaussian
concentration bounds to construct PRS for linear systems subject to additive noise.
Liu \emph{et al.}~\cite{Liu2026probabilistic} developed concentration-based PRS for nonlinear systems with additive noise using the averaged moment-generating function.
These concentration-based approaches can yield less conservative PRS than bounds based solely on the mean and covariance without requiring
the exact state distribution.

In contrast to additive noise, stochastic multiplicative uncertainty presents additional challenges,
since its effect depends on the state. In SMPC, Cannon \emph{et
al.}~\cite{CannonAutomatica2009} introduced invariance with
probability $p$ to address probabilistic constraints and stability,
and subsequently extended their approach to combined multiplicative
and additive stochastic uncertainties~\cite{Cannon2009}.
Fleming and Cannon~\cite{Fleming2018} also addressed this setting
through sample-based approximations of chance constraints.
These contributions are primarily MPC-oriented. Regarding the
propagation of uncertainty, Moussa and Fiacchini~\cite{Moussa2026}
established an exact covariance characterization for systems with
stochastic parametric and additive uncertainties, highlighting the
coupling between the mean and covariance dynamics.
Adu \emph{et al.}~\cite{Adu2026probabilistic} studied sub-Gaussian concentration-based probabilistic reachability for nonlinear systems with non-additive noise. However, their framework does not cover linear multiplicative uncertainty.

In this paper, we extend sub-Gaussian concentration-based techniques to probabilistic reachability for linear systems with bounded stochastic multiplicative uncertainty and potentially unbounded additive noise. The resulting propagation scheme uses a worst-case envelope over the uncertainty support. We also develop a complementary sub-exponential propagation scheme based on expectation-based propagation of the matrix variance proxy, together with a constructive design of the moment-generating function (MGF) domain for affine uncertainty with polytopic support.

Both approaches yield ellipsoidal probabilistic reachable sets and half-space bounds with explicit probability guarantees. For half-space constraints, we establish a necessary and sufficient condition, based on the concentration parameters and the prescribed probability level, under which the sub-exponential margin is no larger than its sub-Gaussian counterpart. Numerical results illustrate how the relative conservatism of the two approaches changes with the probability level.

The remainder of the paper is organized as follows: Section~\ref{sec:problem}
introduces the problem, Sections~\ref{sec:subg}~and~\ref{sec:subexp} develop the propagation
schemes, Section~\ref{sec:prs_bounds} derives the probabilistic bounds, Section~\ref{sec:numerical}
presents numerical results, and Section~\ref{sec:conclusion} concludes the paper.

\textit{Notation:}
For symmetric matrices $ A, B $, $ A \succeq B $ (resp. $A \succ B $) means that $ A - B $ is positive semidefinite (resp. definite).
For a symmetric matrix $S$, $\lambda_{\max}(S)$ and
$\lambda_{\min}(S)$ denote its largest and smallest
eigenvalues, respectively.
For $S \succeq 0$, define $\|x\|_S^2 := x^\top S x$.
For a set $\mathcal{C}$, $\operatorname{conv}(\mathcal{C})$ denotes its convex hull.
The symbols $\mathbb{P}$ and $\mathbb{E}$ denote probability
and expectation, respectively.
For a random vector $X$, $\operatorname{cov}(X)$
denotes its covariance matrix.

\section{Preliminaries and Problem Formulation}
\label{sec:problem}

We consider the discrete-time linear stochastic system
\begin{equation}
    x_{k+1}
    =
    A(\xi_k)x_k+w_k,
    \qquad k\in\mathbb{N},
    \label{eq:system}
\end{equation}
where \(x_k\in\mathbb{R}^n\) denotes the state, 
\(\xi_k\in\mathbb{R}^p\) is a stochastic multiplicative uncertainty, 
and \(w_k\in\mathbb{R}^n\) is an additive process noise. The mapping
\(
    A:\mathbb{R}^p\rightarrow\mathbb{R}^{n\times n}
\)
is assumed measurable. The multiplicative uncertainty vector \(\xi_k\) is assumed to have bounded support.

\begin{assumption}
\label{ass:bounded_xi}
For every \(k\in\mathbb{N}\), there exists a compact set
\(\Xi_k\subset\mathbb{R}^p\) such that
\(\mathbb{P}\{\xi_k\in\Xi_k\}=1.\)
\end{assumption}

No parametric distribution is imposed on \(\xi_k\). In particular, its probability distribution may be non-Gaussian and time-varying. The bounded support will be exploited by the sub-Gaussian propagation developed in Section~\ref{sec:subg}, whereas additional distributional information on \(\xi_k\) will be used in the sub-exponential propagation of Section~\ref{sec:subexp}.

\subsection{Sub-Gaussian and sub-exponential random vectors}
A sub-Gaussian random vector has tail probabilities that decay at least as fast as those of a Gaussian distribution~\cite{Vershynin2018}.
We first recall the matrix-proxy characterization of sub-Gaussian random vectors introduced in~\cite{Ao2025}.
For a random vector $X\in\mathbb{R}^n$, its MGF is defined as $M_X(\lambda):=\mathbb{E}[\exp(\lambda^\top X)]$ for $\lambda \in \bbR^n$, whenever the expectation is finite. The following characterizations provide upper bounds on the MGF of the centered random vector $X-\mu$.

\begin{definition}
\label{def:subg}
An \(\bbR^n\)-valued random vector \(X\) with finite mean
\(\bbE[X]=\mu\in\bbR^n\) is said to be sub-Gaussian with matrix variance proxy
\(\Sigma\succeq0\), denoted by
\(
    X\sim\subg(\mu,\Sigma),
\)
if, for all \(\lambda\in\mathbb{R}^n\), it holds that
\begin{equation}
    M_{X-\mu} (\lambda) \leq
    \exp\!\left(
        \frac{\|\lambda\|_{\Sigma}^2}{2}
    \right).
    \label{eq:mgf_subg}
\end{equation}
\end{definition}

Sub-exponential random vectors can be characterized by MGF bounds that hold in a neighborhood of the
origin~\cite{Spokoiny2013sharp}.
Motivated by the matrix-parameterized 
characterization in Definition~\ref{def:subg}, we use the following sub-exponential characterization, in which both the
MGF bound and its validity domain are parameterized by matrices.

\begin{definition}
\label{def:subexp}
An \(\bbR^n\)-valued random vector \(X\) is said to be sub-exponential with parameters $\mu\in\bbR^n$ and \(\Sigma,S\succeq0\), denoted by
\(
    X\sim\subex(\mu,\Sigma,S),
\)
if, for all \(\lambda\in\bbR^n\) such that \(\|\lambda\|_{S}\leq1\), condition~\eqref{eq:mgf_subg} holds.

If \(X\sim\subex(\mu,\Sigma,0)\), then \(X\) is sub-Gaussian with matrix variance proxy \(\Sigma\), since \(\|\lambda\|_S=0\) for all \(\lambda\in\bbR^n\).
\end{definition}

Sub-exponentiality allows heavier tails than sub-Gaussianity by restricting the domain of the bound on the MGF. As shown later, this relaxation enables an expectation-based treatment of multiplicative uncertainty, potentially reducing the conservatism of worst-case propagation.

The following lemma establishes basic properties of the matrix-parameterized MGF characterization in Definition~\ref{def:subexp}.
\begin{lemma}
\label{prop:subexp_moments}
Suppose that an $\bbR^n$-valued random vector $X$
satisfies $X \sim \subex(\mu, \Sigma, S)$.
Then, we have
\begin{equation}
    \bbE[X]=\mu,
    \qquad
    \cov(X)\preceq\Sigma.
\end{equation}
\end{lemma}

\begin{proof}
Fix any $v\in\bbR^n$ and let
$Z:=v^\top(X-\mu)$ and
$\sigma_v^2 := v^\top\Sigma v$.
For all $t>0$ such that $t^2v^\top Sv\le1$, we have
$
    \bbE[\Ee^{\pm tZ}]
    \le \Ee^{t^2\sigma_v^2/2}
$.
Using the inequality $\Ee^x\ge1+x$, we obtain
$
    \pm\bbE[Z]
    \le
    \frac{1}{t} (\Ee^{t^2\sigma_v^2/2}-1)
$.
Letting $t\downarrow0$ yields
$\bbE[Z]=0$.
Since $v$ is arbitrary, this implies
$\bbE[X]=\mu$.

Next, using $\cosh(x)\ge1+x^2/2$,
we obtain
\begin{align*}
    1+\frac{t^2}{2}\bbE[Z^2]
    \le \frac{\bbE[\Ee^{tZ}]+\bbE[\Ee^{-tZ}]}{2} \le \Ee^{t^2\sigma_v^2/2}.
\end{align*}
Consequently,
$
    \bbE[Z^2]
    \le
    \frac{2}{t^2}
    (\Ee^{t^2\sigma_v^2/2}-1)
$.
Taking $t\downarrow0$ gives
\begin{equation*}
    v^\top\cov(X)v
    =\bbE[Z^2]
    \le v^\top\Sigma v.
\end{equation*}
Since this holds for any $v\in\bbR^n$, we conclude that
$\cov(X)\preceq\Sigma$.
\end{proof}

\subsection{Concentration of the multiplicative uncertainty}

The following assumption provides a matrix variance proxy for the random vector generated by the multiplicative uncertainty acting on a deterministic vector.

\begin{assumption}
\label{ass:A_subgauss}
For every \(k\in\mathbb{N}\), there exists a scalar \(a_k>0\) such that, for all \(v\in\bbR^n\),
\begin{equation}
    A(\xi_k) v \sim \calS \calG \! \left(\bar{A}_k v, a_k \, {\rm cov}(A(\xi_k)v) \right) ,
    \label{eq:A_subgauss}
\end{equation}
where
\(
    \bar A_k:=\bbE[A(\xi_k)]
\)
and
\[
    {\rm cov}(A(\xi_k)v)
    :=
    \bbE\!
    \left[
        \left(
            A(\xi_k)-\bar A_k
        \right)
        vv^\top
        \left(
            A(\xi_k)-\bar A_k
        \right)^\top
    \right].
\]
\end{assumption}

Assumption~\ref{ass:A_subgauss} will be used to bound the uncertainty induced by the term
\(
    (A(\xi_k)-\bar A_k)\mu_k
\) in the propagation of the state distribution, where \( \mu_k \) stands for the state mean. 

The following proposition gives a sufficient condition under which Assumption~\ref{ass:A_subgauss} holds for affine dependence on the multiplicative uncertainty.

\begin{proposition}
\label{prop:A_subgauss_affine}
Let
\(
    A(\xi_k)
    :=
    \mathcal{A}_0
    +
    \sum_{i=1}^{p}
    \mathcal{A}_i \xi_{k,i},
\)
where \(\mathcal{A}_i\in\bbR^{n\times n}\) and \(\xi_{k,i}\in\bbR\) denotes the \(i\)-th component of \(\xi_k\). Assume that
\(
    \xi_k\sim\subg(0,\Sigma_{\xi,k})
\)
and
\(
    \Gamma_{\xi,k}
    :=
    \bbE[\xi_k\xi_k^\top]
    \succ0.
\)
Then, \(A(\xi_k)\) satisfies Assumption~\ref{ass:A_subgauss} with
\(
    a_k
    =
    \lambda_{\max} \!
    \left(
        \Gamma_{\xi,k}^{-1/2}\Sigma_{\xi,k}\Gamma_{\xi,k}^{-1/2}
    \right).
\)
\end{proposition}

The proof is presented in the Appendix.

\subsection{Initial state, additive noise, and independence}

We assume that the initial condition admits the matrix-parameterized MGF characterization introduced above.

\begin{assumption}
\label{ass:initial_condition}
The initial state \(x_0\) satisfies
\(
    x_0\sim\subex(\mu_0,\Sigma_0,S_0),
\)
for some \(\mu_0\in\bbR^n\) and \(\Sigma_0,S_0\succeq0\). The sub-Gaussian case is recovered by taking \(S_0=0\).
\end{assumption}

We assume that the possibly unbounded additive noise is sub-Gaussian.

\begin{assumption}
\label{ass:additive_noise}
For every \(k\in\mathbb{N}\), the additive noise \(w_k\) satisfies
\(
    w_k\sim\subg(0,\Sigma_w),
\)
for some \(\Sigma_w\succeq0\).
\end{assumption}

To state the propagation results in a compact form, we impose the following independence assumption.

\begin{assumption}
\label{ass:one_step_independence}
For every \(k\in\mathbb{N}\), the random vectors \(x_k\), \(\xi_k\), and \(w_k\) are mutually independent.
\end{assumption}

Assumption~\ref{ass:one_step_independence} ensures that the current multiplicative uncertainty \(\xi_k\) can be treated independently from the current state \(x_k\) in the propagation of the MGF bounds. The distribution of \(\xi_k\) may nevertheless be non-Gaussian and time-varying. The additive noise is described by a common matrix variance proxy \(\Sigma_w\), although a time-varying proxy \(\Sigma_{w,k}\) could be considered without changing the structure of the propagation rules.

\subsection{Problem statement  (probabilistic reachability)}
Since the initial state is assumed to be random, we consider the following unconditional formulation of \(k\)-step probabilistic reachability, adapted from \cite{Hewing2020}, for a prescribed violation probability \(\delta\in(0,1)\).

\begin{definition}[\(k\)-step probabilistic reachable set]
\label{def:k_step_prs}
A measurable set \(\mathcal{R}_k^\delta\subseteq\mathbb{R}^n\) is called a \(k\)-step probabilistic reachable set, or \(k\)-step PRS, with probability level \(1-\delta\) for system~\eqref{eq:system} if
\begin{equation}
    \mathbb{P} \!
    \left\{
        x_k\in\mathcal{R}_k^\delta
    \right\}
    \geq
    1-\delta.
    \label{eq:k_step_prs_def}
\end{equation}
\end{definition}

In this paper, we address the construction of $k$-step
probabilistic reachable sets over a finite horizon $N$.
More specifically, our objective is to derive deterministic
recursions characterizing the concentration properties
of the state distribution, and exploit them to construct
sets $\mathcal{R}_k^\delta$ satisfying~\eqref{eq:k_step_prs_def} for $k = 0,\ldots,N$.

We consider both ellipsoidal confidence regions and
half-space bounds, the latter being particularly relevant
for the enforcement of linear chance constraints.

To this end, Sections~\ref{sec:subg} and~\ref{sec:subexp}
develop two complementary propagation schemes based,
respectively, on sub-Gaussian and sub-exponential
matrix-parameterized MGF characterizations.

\section{Sub-Gaussian Propagation}
\label{sec:subg}

We first derive a propagation rule based on the sub-Gaussian characterization of Definition~\ref{def:subg}. In this case, the effect of the multiplicative uncertainty on the matrix variance proxy is handled through a worst-case envelope over the support \(\Xi_k\).

\begin{proposition}[Sub-Gaussian propagation]
\label{prop:subg_propagation}
Let \(k\in\mathbb{N}\) be fixed, and let Assumptions~\ref{ass:bounded_xi}, 
\ref{ass:A_subgauss}, \ref{ass:additive_noise}, and~\ref{ass:one_step_independence} hold. 
Assume that
\(
    x_k\sim\subg(\mu_k,\Sigma_k),
\)
and let \(\bar\Sigma_k\succeq0\) be such that
\begin{equation}
    A(\xi)\Sigma_kA(\xi)^\top
    \preceq
    \bar\Sigma_k,
    \qquad
    \forall \xi\in\Xi_k.
    \label{eq:subg_worst_case_envelope}
\end{equation}
Then
\[  x_{k+1}\sim\subg(\mu_{k+1},\Sigma_{k+1}),
\]
where
\begin{align}
\mu_{k+1} &= \bar A_k\mu_k,
\label{eq:subg_mean_recursion}\\
\Sigma_{k+1} &= \bar\Sigma_k
+ a_k\,{\rm cov}(A(\xi_k)\mu_k)
+ \Sigma_w.
\label{eq:subg_proxy_recursion}
\end{align}
\end{proposition}

\begin{proof}
By \eqref{eq:system} and \eqref{eq:subg_mean_recursion}, one has
\[
    x_{k+1}-\mu_{k+1}
    =
    A(\xi_k)(x_k-\mu_k)
    +
    \left(
        A(\xi_k)-\bar A_k
    \right)\mu_k
    +
    w_k.
\]
Let \(\lambda\in\bbR^n\). Using Assumption~\ref{ass:one_step_independence} and conditioning on \(\xi_k\), we obtain
\[
\begin{aligned}
    &\bbE
    \! \left[
        \exp\!\left(
            \lambda^\top
            A(\xi_k)(x_k-\mu_k)
        \right)
        \,\middle|\,
        \xi_k
    \right]
    \\
    &\hspace{1cm}
    \leq
    \exp\!\left(
        \frac{1}{2}
        \lambda^\top
        A(\xi_k)\Sigma_kA(\xi_k)^\top
        \lambda
    \right)
    \\
    &\hspace{1cm}
    \leq
    \exp\!\left(
        \frac{1}{2}
        \|\lambda\|_{\bar\Sigma_k}^2
    \right),
\end{aligned}
\]
where the last inequality follows from \eqref{eq:subg_worst_case_envelope}. Moreover, Assumption~\ref{ass:A_subgauss} gives
\[
    \bbE
    \! \left[
        \exp\!\left(
            \lambda^\top
            \! \left(
                A(\xi_k)-\bar A_k
            \right)\mu_k
        \right)
    \right]
    \leq
    \exp\!\left(
        \frac{1}{2}
        \|\lambda\|^2_{
            a_k\,{\rm cov}(A(\xi_k)\mu_k)
        }
    \right).
\]
Finally, since \(w_k\sim\subg(0,\Sigma_w)\), the additive noise contributes the proxy \(\Sigma_w\). Combining the above bounds through the tower property gives
\[
\begin{aligned}
    &\bbE
    \! \left[
        \exp\!\left(
            \lambda^\top(x_{k+1}-\mu_{k+1})
        \right)
    \right]
    \\
    &=
    \bbE
    \Bigl[
        \exp\!\left(
            \lambda^\top
            \!\left(
                A(\xi_k)-\bar A_k
            \right)\mu_k
        \right)
        \\
        &\quad \times \bbE
        \! \left[
            \exp\!\left(
                \lambda^\top
                A(\xi_k)(x_k-\mu_k)
            \right)
            \,\middle|\,
            \xi_k
        \right]
    \Bigr]
    \bbE
    \! \left[
        \exp\!\left(
            \lambda^\top w_k
        \right)
    \right]
    \\
    &\leq
    \exp\!\left(
        \frac{1}{2}
        \|\lambda\|_{\bar\Sigma_k}^2
    \right)
    \exp\!\left(
        \frac{1}{2}
        \|\lambda\|^2_{
            a_k\,{\rm cov}(A(\xi_k)\mu_k)
        }
    \right)
    \\
    &\quad \times \exp\!\left(
        \frac{1}{2}
        \|\lambda\|_{\Sigma_w}^2
    \right)
    \\
    &=
    \exp\!\left(
        \frac{1}{2}
        \|\lambda\|_{\Sigma_{k+1}}^2
    \right),
\end{aligned}
\]
with \(\Sigma_{k+1}\) given by \eqref{eq:subg_proxy_recursion}. Hence,
\(x_{k+1}\sim\subg(\mu_{k+1},\Sigma_{k+1})\).
\end{proof}
\begin{remark}\label{rmk:vertices}
A matrix $\bar\Sigma_k$ satisfying
\eqref{eq:subg_worst_case_envelope} exists, in particular,
when $A(\cdot)$ is continuous on the compact support
$\Xi_k$.
Moreover, when $\Xi_k$ is polytopic and $A(\xi)$ is affine in $\xi$ as in Proposition~\ref{prop:robust_S_design}, condition~\eqref{eq:subg_worst_case_envelope} can be verified by checking only the vertices of $\Xi_k$.
\hfill $\blacklozenge$
\end{remark}

Starting from an initial condition satisfying
\(x_0\sim\subg(\mu_0,\Sigma_0)\), as stated in
Assumption~\ref{ass:initial_condition} for \(S_0=0\),
Proposition~\ref{prop:subg_propagation} can be applied
recursively to obtain a sub-Gaussian characterization
of \(x_k\) for all \(k\in\mathbb N\).
The matrix \(\bar\Sigma_k\) in
\eqref{eq:subg_worst_case_envelope} provides a worst-case
upper bound over the support \(\Xi_k\) of the multiplicative
uncertainty, whereas the mean-induced term and its
associated coefficient \(a_k\) exploit distributional
information on this uncertainty.

\section{Sub-Exponential Propagation}
\label{sec:subexp}

We now derive a complementary propagation rule based on the sub-exponential characterization of Definition~\ref{def:subexp}. In contrast with the sub-Gaussian construction, the restricted MGF domain allows the propagation of the matrix variance proxy to involve the expectation
\(
    \bbE[A(\xi_k)\Sigma_k A(\xi_k)^\top]
\),
rather than only a worst-case envelope over \(\Xi_k\).
This expectation term has the same structure as that appearing in the exact covariance propagation under multiplicative uncertainty~\cite{Moussa2026}.

\begin{theorem}[Sub-exponential propagation]
\label{prop:subexp_propagation}
Let \(k\in\mathbb{N}\) be fixed, and let Assumptions~\ref{ass:bounded_xi}, \ref{ass:A_subgauss}, \ref{ass:additive_noise}, and~\ref{ass:one_step_independence} hold. Assume that
\(
    x_k\sim\subex(\mu_k,\Sigma_k,S_k),
\)
with \(S_k\succ0\). Let \(S_{k+1}\succ0\) satisfy
\begin{equation}
    S_{k+1}
    \succeq
    A(\xi)S_kA(\xi)^\top,
    \qquad
    \forall \xi\in\Xi_k.
    \label{eq:subexp_S_propagation}
\end{equation}
Define
\begin{equation}
    r_k
    :=
    \lambda_{\max}
    \! \left(
        S_k^{-1/2 }\Sigma_k S_k^{-1/2}
    \right),
    \qquad
    \rho_k
    :=
    \frac{\exp(r_k)-1}{r_k},
    \label{eq:subexp_rho}
\end{equation}
with \(\rho_k=1\) when \(r_k=0\). Then
\begin{equation}\label{eq:subex_one_step}
    x_{k+1}
    \sim
    \subex(\mu_{k+1},\Sigma_{k+1},S_{k+1}),
\end{equation}
where
\begin{align}
\mu_{k+1} &= \bar A_k\mu_k,
\label{eq:subexp_mean_recursion}\\
\Sigma_{k+1} &= \rho_k
    \bbE
    \! \left[
        A(\xi_k)\Sigma_kA(\xi_k)^\top
    \right]
    +
    2a_k\,{\rm cov}(A(\xi_k)\mu_k)
    +
    \Sigma_w .
    \label{eq:subexp_proxy_recursion}
\end{align}
\end{theorem}

\begin{proof}
Let
$
    \Lambda_k
    :=
    \left\{
        \lambda\in\bbR^n:
        \|\lambda\|_{S_k}\leq1
    \right\}
$.
We show that the MGF bound in Definition~\ref{def:subexp} holds at time \(k+1\) for all \(\lambda\in\Lambda_{k+1}\). From \eqref{eq:system} and \eqref{eq:subexp_mean_recursion}, one has
\[
    x_{k+1}-\mu_{k+1}
    =
    A(\xi_k)(x_k-\mu_k)
    +
    \left(
        A(\xi_k)-\bar A_k
    \right)\mu_k
    +
    w_k .
\]
Let \(\lambda\in\Lambda_{k+1}\). By \eqref{eq:subexp_S_propagation},
\[
    \|A(\xi)^\top\lambda\|_{S_k}^2
    =
    \|\lambda\|_{A(\xi)S_kA(\xi)^\top}^2
    \leq
    \|\lambda\|_{S_{k+1}}^2
    \leq1,
    \quad
    \forall \xi\in\Xi_k.
\]
Hence, since \(x_k\) is independent of \(\xi_k\) from Assumption~\ref{ass:one_step_independence}, the sub-exponential
MGF bound for \(x_k\) can be applied conditionally on \(\xi_k\), yielding
\[
    \bbE
    \! \left[
        \exp\!\left(
            \lambda^\top
            A(\xi_k)(x_k-\mu_k)
        \right)
        \,\middle|\,
        \xi_k
    \right]
    \leq
    \exp\!\left(
        \frac{1}{2}
        \|\lambda\|^2_{A(\xi_k)\Sigma_kA(\xi_k)^\top}
    \! \right) \! .
\]
Therefore,
\[
\begin{aligned}
    &\bbE
    \! \left[
        \exp\!\left(
            \lambda^\top(x_{k+1}-\mu_{k+1})
        \right)
    \right]
    \\
    &\leq
    \exp\!\left(
        \frac{1}{2}
        \|\lambda\|_{\Sigma_w}^2
    \right)\\
    &\quad \times
    \bbE
    \! \left[
        \exp\!\left(
            \lambda^\top
            \left(
                A(\xi_k)-\bar A_k
            \right)\mu_k
            +
            \frac{1}{2}
            \|\lambda\|^2_{A(\xi_k)\Sigma_kA(\xi_k)^\top}
        \right)
    \right].
\end{aligned}
\]
Applying the Cauchy--Schwarz inequality to the last expectation gives
\begin{align}
    &\bbE
    \! \left[
        \exp\!\left(
            \lambda^\top(x_{k+1}-\mu_{k+1})
        \right)
    \right]
    \nonumber\\
    &\leq
    \exp\!\left(
        \frac{1}{2}
        \|\lambda\|_{\Sigma_w}^2
    \right)
    \bbE
    \! \left[
        \exp\!\left(
            2\lambda^\top
            \left(
                A(\xi_k)-\bar A_k
            \right)\mu_k
        \right)
    \right]^{1/2}
    \nonumber\\
    &\quad\times
    \bbE
    \! \left[
        \exp\!\left(
            \|\lambda\|^2_{A(\xi_k)\Sigma_kA(\xi_k)^\top}
        \right)
    \right]^{1/2}.  \label{eq:MGF_next_bound}
\end{align}
By Assumption~\ref{ass:A_subgauss},
\[
\begin{aligned}
    &\bbE
    \! \left[
        \exp\!\left(
            2\lambda^\top
            \left(
                A(\xi_k)-\bar A_k
            \right)\mu_k
        \right)
    \right]^{1/2}
    \\
    & \qquad \leq
    \exp\!\left(
        \frac{1}{2}
        \|\lambda\|^2_{
            2a_k\,{\rm cov}(A(\xi_k)\mu_k)
        }
    \right).
\end{aligned}
\]
It remains to bound the last factor in \eqref{eq:MGF_next_bound}. Define
\[
    y_\xi(\lambda)
    :=
    \|\lambda\|^2_{A(\xi)\Sigma_kA(\xi)^\top}.
\]
Since
$
    \Sigma_k
    \preceq
    r_k S_k
$,
we have, for all \(\xi\in\Xi_k\),
\[
    y_\xi(\lambda)
    \leq
    r_k
    \|\lambda\|^2_{A(\xi)S_kA(\xi)^\top}
    \leq
    r_k
    \|\lambda\|^2_{S_{k+1}}
    \leq
    r_k .
\]
Using the inequality
\[
    \exp(y)
    \leq
    1
    +
    \frac{\exp(r_k)-1}{r_k}y,
    \qquad
    y\in[0,r_k],
\]
and \(1+y\leq\exp(y)\), we obtain
\[
    \bbE
    \!\left[
        \exp(y_{\xi_k}(\lambda))
    \right]
    \leq
    \exp
    \! \left(
        \rho_k
        \bbE
        \! \left[
            y_{\xi_k}(\lambda)
        \right]
    \right).
\]
Thus,
\[
\begin{aligned}
     &\bbE
    \! \left[
        \exp
        \left(
            \|\lambda\|^2_{A(\xi_k)\Sigma_kA(\xi_k)^\top}
        \right)
    \right]^{1/2}\\
   & \qquad \leq
    \exp\!\left(
        \frac{1}{2}
        \|\lambda\|^2_{
            \rho_k
            \bbE[A(\xi_k)\Sigma_kA(\xi_k)^\top]
        }
    \right). 
\end{aligned}
\]
Combining the previous inequalities gives
\[
    \bbE
    \! \left[
        \exp\!\left(
            \lambda^\top(x_{k+1}-\mu_{k+1})
        \right)
    \right]
    \leq
    \exp\!\left(
        \frac{1}{2}
        \|\lambda\|_{\Sigma_{k+1}}^2
    \right),
\]
where \(\Sigma_{k+1}\) is given by \eqref{eq:subexp_proxy_recursion}. Since the inequality holds for all \(\lambda\in\Lambda_{k+1}\), this proves \eqref{eq:subex_one_step}.
\end{proof}

Under Assumption~\ref{ass:initial_condition} with $S_0 \succ 0$, Theorem~\ref{prop:subexp_propagation} can be applied recursively to obtain a sub-exponential characterization of $x_k$ at each time step.

Theorem~\ref{prop:subexp_propagation} highlights the distribution-sensitive nature of the sub-exponential propagation. The term
\(
    \bbE
    \! \left[
        A(\xi_k)\Sigma_kA(\xi_k)^\top
    \right]
\)
depends on the probability distribution of the multiplicative uncertainty, whereas the matrix \(S_{k+1}\) is propagated through the robust condition~\eqref{eq:subexp_S_propagation}. Hence, the sub-exponential construction combines an expectation-based variance proxy with a worst-case control of the MGF domain, potentially reducing the conservatism associated with the support-based propagation in Proposition~\ref{prop:subg_propagation}.

Condition~\eqref{eq:subexp_S_propagation} requires the selection of a matrix $S_{k+1}$ that preserves the MGF domain under every admissible realization of the multiplicative uncertainty. For the affine dependence considered in Proposition~\ref{prop:A_subgauss_affine}, and when the uncertainty support is polytopic, this condition can be verified at the vertices. The following proposition exploits this property to provide a constructive procedure for selecting $S_{k+1}$ while controlling the coefficient entering the subsequent propagation.

\begin{proposition}[Design of $S_{k+1}$]\label{prop:robust_S_design}
Let $\xi\in\mathbb{R}^p$ be a random vector with support contained in the polytope
\[
\Xi:=\mathrm{conv}\{\xi^{(1)},\dots,\xi^{(L)}\},\qquad \bbP\{\xi\in\Xi\}=1,
\]
and assume that $A(\xi)$ depends affinely on $\xi$. Define the vertex matrices
$A^{(i)}:=A(\xi^{(i)})$, $i=1,\dots,L$.
Let $\Sigma\succeq 0$, $S_k\succ 0$, and $\nu > 0$ be given.
Consider the decision variables $S\succ 0$ and $t> 0$ and the optimization problem
\begin{align}
\min_{S,t}\quad & t+\nu\,\mathrm{trace}(S) \label{eq:sdp_cost}\\
\text{s.t.}\quad & \Sigma \preceq t\,S, \label{eq:lmi_spectral}\\
& S \succeq A^{(i)} S_k A^{(i)\top}, \qquad i=1,\dots,L. \label{eq:lmi_vertices}
\end{align}
Assume that an optimal solution $(S^\star,t^\star)$ exists.
Then, choosing $S_{k+1}=S^\star$ satisfies \eqref{eq:subexp_S_propagation}, and
\[
\lambda_{\max} \big((S^\star)^{-1/2} \Sigma (S^\star)^{-1/2}\big)\le t^\star.
\]
In particular, when $\Sigma=\Sigma_{k+1}$, the coefficient
\[
\widehat\rho_{k+1} := \frac{\exp(t^\star)-1}{t^\star}
\]
is an admissible upper bound on $\rho_{k+1}$ in \eqref{eq:subexp_rho},
and can be used conservatively in the subsequent update \eqref{eq:subexp_proxy_recursion}.
\end{proposition}

\begin{proof}
Since $A(\xi)$ depends affinely on $\xi$, each $\xi\in\Xi$ admits weights
$\theta_i\ge0$, $\sum_i\theta_i=1$, such that
$A(\xi)=\sum_i\theta_i A^{(i)}$. Therefore,
\[
A(\xi)S_k A(\xi)^\top
\preceq \sum_{i=1}^L\theta_i A^{(i)}S_k A^{(i)\top}
\preceq S^\star,
\]
where the first inequality follows by convexity and the second by
\eqref{eq:lmi_vertices}. Thus $S_{k+1}=S^\star$ satisfies
\eqref{eq:subexp_S_propagation}.

Define $\rho(t):=(\exp(t)-1)/t$ for $t>0$, with $\rho(0)=1$.
The function $\rho(\cdot)$ is nondecreasing, and $\rho(t)>1$ for $t>0$.
From \eqref{eq:lmi_spectral} and $S^\star\succ0$, we have
$(S^\star)^{-1/2}\Sigma(S^\star)^{-1/2}\preceq t^\star I$,
and therefore
$\lambda_{\max} \big((S^\star)^{-1/2}\Sigma(S^\star)^{-1/2}\big)\le t^\star$.
For $\Sigma=\Sigma_{k+1}$, monotonicity of $\rho$ yields
\[
\rho_{k+1}\le \rho(t^\star)
=\frac{\exp(t^\star)-1}{t^\star}=\widehat\rho_{k+1}.
\]
Hence, $\widehat\rho_{k+1}$ may replace $\rho_{k+1}$ as a conservative
coefficient in the following variance-proxy recursion.
Moreover, the objective \eqref{eq:sdp_cost} promotes small $t$ (and thus a
coefficient close to $1$) while regularizing the size of $S$ through
$\nu\,\mathrm{trace}(S)$.
\end{proof}

For fixed $t>0$, the constraints in
\eqref{eq:lmi_spectral} and \eqref{eq:lmi_vertices} are linear matrix
inequalities in $S$. Since the constraint $\Sigma\preceq tS$ is not
jointly linear in $(S,t)$, the design can be implemented using a scalar
search over $t$, solving a semidefinite program for each fixed value.
The parameter $\nu>0$ balances a smaller bound on $\rho_{k+1}$ against a larger MGF domain. In practice, $\nu$ can be tuned to reduce the resulting probability bounds.

\section{Probabilistic Reachability Bounds}
\label{sec:prs_bounds}

The propagation rules in Sections~\ref{sec:subg} and~\ref{sec:subexp}
provide matrix-parameterized MGF characterizations of the state distribution.
We now use these characterizations to construct probabilistic bounds
with an explicit confidence level. We first consider ellipsoidal
regions and then half-space bounds, the latter being directly suited
to linear chance constraints.

\subsection{Ellipsoidal probabilistic reachable sets}

For the sub-Gaussian characterization, we use the ellipsoidal confidence bound of~\cite{Ao2025} to construct \(k\)-step probabilistic reachable sets. Let
$
    g(s)
    :=
    \frac{\Ee^s}{1+s}
$ for $s\geq0$.
For a prescribed violation probability \(\delta\in(0,1)\), define
\begin{equation}
    q_{\rm sg}(\delta)
    :=
    n
    +
    n\,g^{-1}
    \left(
        \delta^{-2/n}
    \right).
    \label{eq:subg_gamma}
\end{equation}

\begin{lemma}[Sub-Gaussian ellipsoidal \(k\)-step PRS \cite{Ao2025}]
\label{cor:subg_prs}
Assume that \(x_k\sim\subg(\mu_k,\Sigma_k)\) with \(\Sigma_k\succ0\). Then the set
\begin{equation}
    \mathcal{E}_{k,{\rm sg}}^\delta
    :=
    \left\{
        x\in\bbR^n:
        \|x-\mu_k\|_{\Sigma_k^{-1}}^2
        \leq
        q_{\rm sg}(\delta)
    \right\}
    \label{eq:subg_prs}
\end{equation}
is a \(k\)-step PRS with probability level \(1-\delta\), namely
\[
    \bbP
    \! \left\{
        x_k\in\mathcal{E}_{k,{\rm sg}}^\delta
    \right\}
    \geq
    1-\delta.
\]
\end{lemma}

The restricted MGF domain of the sub-exponential characterization
leads to a two-regime confidence bound: a quadratic-exponential regime
and a linear-exponential regime. The following proposition establishes
the corresponding ellipsoidal tail bound.

\begin{proposition}[Sub-exponential ellipsoidal bound]\label{prop:elliptical_bound} For $ X \sim \subex(\mu,\Sigma, \Sigma/d^2) $ with $ \Sigma \succ 0 $ and $d>0$, it holds for any $\tau \ge 0$,
\begin{align} 
&\bbP \! \left\{ \| X - \mu \|_{\Sigma^{-1}} \ge \tau \right\} \nonumber\\
&\le 
\begin{cases} 
5^n \exp\! \left( - \frac{\tau^2}{8} \right), & 0 \le \tau \le 2d, \\ 
5^n \exp\! \left( - \frac{d\tau}{2} + \frac{d^2}{2} \right), & \tau > 2d . 
\end{cases}  \label{eq:subex_ellipsoidal}
\end{align} 
\end{proposition}

\begin{proof}
Let $ Z := \Sigma^{-1/2} (X - \mu) $. Then, $ X \sim \subex(\mu,\Sigma, \Sigma/d^2) $ means
\begin{align}
    \bbE[ \exp(v^\top Z) ] \le \exp \! \left( \frac{\|v\|^2}{2}  \right) , \ \forall v\in \bbR^n \ {\rm s.t.} \ \| v \|^2 \le d^2 . \label{eq:sub_exp_equiv}
\end{align}
Let $ \calS^{n-1} $ be the unit sphere in $ \bbR^n $. 
Since $\|Z\|=\sup_{u\in\calS^{n-1}}u^\top Z$,
we use an $\varepsilon$-net~\cite{Vershynin2018} to bound the norm
through finitely many directional projections.

By the Chernoff bound, for any $ u \in \calS^{n-1} $, $ s > 0 $, and $ \tau \in \bbR $, it holds that
\begin{align*}
    \bbP \! \left\{ u^\top Z \ge \tau \right\} \le \exp(-s\tau) \bbE\! \left[ \exp(su^\top Z) \right] .
\end{align*}
Moreover, $ v := s u $ with $ u\in \calS^{n-1} $ and $ s > 0 $ satisfies \eqref{eq:sub_exp_equiv} as long as $ \| v \| = s \le d $. That is,
\begin{equation*}
    \bbE[ \exp(su^\top Z) ] \le \exp \! \left( \frac{s^2}{2}   \right) , \ \forall u\in \calS^{n-1}, \ \forall s \in (0,d] .
\end{equation*}
Therefore, for any $ u\in \calS^{n-1} $, we have
\begin{align}
    \bbP\! \left\{ u^\top Z \ge \tau  \right\} &\le \inf_{s\in (0,d]} \exp\! \left( -s \tau + \frac{s^2}{2}  \right) \nonumber\\
    &= \begin{cases}
        \exp\! \left( - \frac{\tau^2}{2} \right) , & \tau \in [0,d], \\
        \exp\! \left( -d \tau + \frac{d^2}{2} \right), & \tau > d .
    \end{cases}\label{eq:uZ_bound}
\end{align}

Let $ \calN_{\varepsilon} $ be an $ \varepsilon $-net of $ \calS^{n-1} $, i.e., a finite subset such that every point of
$\calS^{n-1}$ is within Euclidean distance
$\varepsilon$ of some point in $\calN_\varepsilon$.
By a standard covering-number bound
\cite[Corollary~4.2.13]{Vershynin2018},
we can choose $ \calN_{1/2}$ such that
\begin{equation}
    \| Z \| \le 2 \sup_{u\in \calN_{1/2}} u^\top Z, \qquad  |\calN_{1/2}| \le 5^n .
\end{equation}
Thus, it follows that
\begin{align}
    \bbP\! \left\{ \| Z \| \ge \tau  \right\} &\le \bbP\! \left\{ \sup_{u\in \calN_{1/2}} u^\top Z \ge \frac{\tau}{2} \right\} \nonumber\\
    &\le \sum_{u\in \calN_{1/2}} \bbP \! \left\{ u^\top Z \ge \frac{\tau}{2}  \right\} \nonumber\\
    &\le 5^n \max_{u\in \calN_{1/2}} \bbP\! \left\{ u^\top Z \ge \frac{\tau}{2} \right\} . 
\end{align}
By combining this with \eqref{eq:uZ_bound}, we obtain the desired result.
\end{proof}

Proposition~\ref{prop:elliptical_bound} can be applied to $ \subex(\mu,\Sigma,S) $ with
$\Sigma\succ0$ and $S\succ0$ because $ X \sim \subex(\mu,\Sigma,S) $ satisfies $ X \sim \subex(\mu,\Sigma,\Sigma/\lambda_{\min}(S^{-1/2} \Sigma S^{-1/2})) $.

The factor $5^n$ in \eqref{eq:subex_ellipsoidal} arises from the covering-number bound
for the $1/2$-net and the subsequent union bound, which may introduce
conservatism by not exploiting the dependence among
directional projections. 
However, as shown in the following corollary, this factor enters the confidence radius only through $n \log 5$.

\begin{corollary}[Sub-exponential ellipsoidal confidence region]\label{cor:elliptical_confidence}
Let $X\sim\subex(\mu,\Sigma,\Sigma/d^2)$ with $\Sigma\succ 0$ and $d>0$. 
For any $\delta\in(0,1)$, it holds that $\bbP\{X\in\calE\}\ge 1-\delta$, where
\[
\calE(\mu,\Sigma,n,d,\delta)
:=\left\{x\in\mathbb{R}^n:\ \|x-\mu\|_{\Sigma^{-1}}^2 \le \tau^2(n,d,\delta)\right\},
\]
and with $c_{n,\delta}:=n\log 5+\log \delta^{-1}$,
\[
\tau(n,d,\delta) :=
\begin{cases}
d+\dfrac{2c_{n,\delta}}{d}, & d^2 < 2c_{n,\delta},\\[6pt]
\sqrt{8c_{n,\delta}}, & d^2 \ge 2c_{n,\delta}.
\end{cases}
\]
\end{corollary}

\begin{proof}
The proof follows from assigning the tail probabilities in Proposition~\ref{prop:elliptical_bound} to $\delta$. Considering $\tau > 2d$, we have:
\[
    5^n \exp \left(-\frac{d \tau}{2}+ \frac{d^2}{2} \right) = \delta,
\]
and then:
\[
\tau = 2 \frac{n}{d} \log 5 + \frac{2}{d} \log  \delta^{-1} + d.
\]
From $\tau > 2d$, the derived bound holds for $ d^2 < 2c_{n,\delta}$. 

Moreover, for $\tau \le 2d$, the bound can be derived from:
\[
5^n \exp\left(-\frac{\tau^2}{8}\right) = \delta,
\]
leading to $\tau^2 = 8n \log 5 + 8 \log \delta^{-1}$. Finally, one can notice the continuity of the two bounds and that the latter holds for $ d^2 \ge 2 c_{n,\delta}$.
\end{proof}

Applying Corollary~\ref{cor:elliptical_confidence} with the propagated
parameters $(\mu_k,\Sigma_k,S_k)$ yields an ellipsoidal $k$-step PRS.
In particular, for $\Sigma_k\succ0$ and $S_k\succ0$, one may take
$d_k^2=\lambda_{\min}(S_k^{-1/2}\Sigma_k S_k^{-1/2})$ in the
confidence factor.

\subsection{Half-space bounds and comparison}

For a linear chance constraint, the relevant tail probability concerns
the scalar projection $h^\top(X-\mu)$, where $h\in \bbR^n $ defines the
constraint direction. We therefore consider directional concentration
bounds associated with the two characterizations, starting with the
sub-exponential case.

\begin{proposition}[Sub-exponential half-space bound]\label{prop:half_space_bound}
Let $X\in\mathbb{R}^n$ satisfy $ X \sim \subex (\mu, \Sigma, S) $. 
For $h\in\mathbb{R}^n$, define
\begin{equation}
\label{eq:bernstein_vs}
v_h:=h^\top\Sigma h,
\qquad
s_h:=\sqrt{h^\top S h}.
\end{equation}
Then, for any $\delta\in(0,1)$,
\begin{equation}
\label{eq:bernstein_probability}
\mathbb{P}\! \left\{
h^\top(X-\mu)
>
m_{\mathrm{SE}}(v_h,s_h,\delta)
\right\}
\le \delta,
\end{equation}
where
\begin{align}
\label{eq:bernstein_margin}
&m_{\mathrm{SE}}(v,s,\delta)\nonumber
\\
&:=
\begin{cases}
\sqrt{2v\log(1/\delta)},
&
s=0,
\\
\sqrt{2v\log(1/\delta)},
&
s>0
\ \text{and}\
\log(1/\delta)\le\frac{v}{2s^2},
\\
s\log(1/\delta)+\frac{v}{2s},
&
s>0
\ \text{and}\
\log(1/\delta)>\frac{v}{2s^2}.
\end{cases}
\end{align}
\end{proposition}

\begin{proof}
We first consider the case $ s_h > 0 $.
For any $t\in[0,1/s_h]$ and $ a > 0 $, the Chernoff bound and the definition of
sub-exponentiality give
\begin{align}
\mathbb{P}\! \left\{h^\top(X-\mu)>a\right\}
&\le \exp(-ta) \bbE \!  \left[ \exp(t h^\top (X-\mu)) \right] \nonumber\\
&\le \exp\! \left(-ta+\frac{1}{2}t^2v_h\right). \label{eq:half_sp_bound1}
\end{align}
If $ v_h > 0 $ and $a\le v_h/s_h$, the right-hand side of \eqref{eq:half_sp_bound1} is minimized over $ t \in [0,1/s_h] $ at $t=a/v_h$, and thus, we obtain
\begin{equation}\label{eq:half_sp_bound2}
\mathbb{P}\! \left\{h^\top(X-\mu)>a\right\}
\le
\exp\! \left(-\frac{a^2}{2v_h}\right).
\end{equation}
Setting $a=\sqrt{2v_h \log(1/\delta)}$ gives the desired bound \eqref{eq:bernstein_probability}, where the condition $a\le v_h/s_h$ is equivalent to $ \log(1/\delta)\le\frac{v_h}{2s_h^2} $. 

Otherwise, the minimum of \eqref{eq:half_sp_bound1} over $ t \in [0,1/s_h] $ is attained at
$t=1/s_h$, which yields
\[
\mathbb{P}\! \left\{h^\top(X-\mu)>a\right\}
\le
\exp\! \left(
-\frac{a}{s_h}+\frac{v_h}{2s_h^2}
\right).
\]
Taking $a=s_h\log (1/\delta )+\frac{v_h}{2s_h}$ with $ \log (1/\delta )>\frac{v_h}{2s_h^2} $, we obtain the desired bound \eqref{eq:bernstein_probability} for the case where $ s_h > 0 $ and $ \log (1/\delta )>\frac{v_h}{2s_h^2} $.

We next consider the case $s_h=0$. Then, \eqref{eq:half_sp_bound1} is valid for any $t>0$. Consequently, if $ v_h > 0 $, \eqref{eq:half_sp_bound2} holds for any $ a > 0 $. Setting $ a = \sqrt{2 v_h \log (1/\delta)} $ gives the desired bound~\eqref{eq:bernstein_probability}.
If $v_h=0$, \eqref{eq:half_sp_bound1} reduces to \[ \mathbb{P}\!\left\{h^\top(X-\mu)>a\right\}\le \Ee^{-ta} \] for any $a>0$ and $t>0$. Letting $t\to\infty$ and then $a\downarrow0$ gives \[ \mathbb{P}\!\left\{h^\top(X-\mu)>0\right\}=0, \] which proves \eqref{eq:bernstein_probability}.
This completes the proof.
\end{proof}

Setting $ S = 0 $ in Proposition~\ref{prop:half_space_bound}, we recover the sub-Gaussian half-space bound in \cite[Lemma~2]{Ao2025}.
Specifically, $ X \sim \subg (\mu,\Sigma) $ satisfies 
the following probability bound for any $ \delta \in (0,1) $:
\begin{equation}
\label{eq:subgauss_half_space}
\mathbb{P}\! \left\{
h^\top(X-\mu)
>
m_{\mathrm{SG}}(v_h,\delta)
\right\}
\le \delta,
\end{equation}
where
\[
m_{\rm SG}(v,\delta)
:=
\sqrt{2v\log(1/\delta)}.
\]
When both characterizations are available for the same
state $x_k$, we distinguish their respective parameters
by $\Sigma_{G,k}$ and $(\Sigma_{E,k},S_{E,k})$.
For simplicity, the time index is omitted in the
following comparison.
Either characterization can be used to enforce a
given half-space chance constraint.
The following proposition provides a necessary and
sufficient condition under which the sub-exponential
margin is no larger than its sub-Gaussian counterpart.

\begin{proposition}[Sub-Gaussian/sub-exponential dominance]
\label{prop:dominance}
Suppose that $X\in \bbR^n$ satisfies
$X\sim\subg(\mu,\Sigma_G)$ and $X\sim\subex(\mu,\Sigma_E,S_E)$.
For $h\in\mathbb{R}^n$, define
\[
v_G:=h^\top\Sigma_G h,\qquad
v_E:=h^\top\Sigma_E h,\qquad
s_E:=\sqrt{h^\top S_Eh}.
\]
Suppose that $ s_E > 0 $ and $ v_E > 0 $, and let
\[
d:=\frac{\sqrt{v_E}}{s_E},\qquad
\gamma:=\frac{v_G}{v_E}.
\]
Then, for any $ \delta \in (0,1) $, $ m_{\rm SE} (v_E, s_E, \delta) \le m_{\rm SG} (v_G, \delta) $ if and only if
\begin{equation}
    \gamma \ge \Gamma (d,\delta) :=
    \begin{cases}
        1, & \log (1/\delta) \le d^2 /2 , \\
        \frac{(d^{-1}\log (1/\delta) + \frac{d}{2})^2}{2 \log (1/\delta)}, & \log (1/\delta) > d^2 /2 .
    \end{cases}
\end{equation}
\end{proposition}

\begin{proof}
Let $ L:=\log(1/\delta) $. We first consider the case
$
L\le\frac{d^2}{2}
$.
In this case,
\[
m_{\rm SE}(v_E,s_E,\delta)
=
\sqrt{2v_E L},
\]
while
\begin{equation}\label{eq:m_SG1}
m_{\rm SG}(v_G,\delta)
=
\sqrt{2v_G L}
=
\sqrt{v_E}\sqrt{2\gamma L}
\end{equation}
holds for any $ (v_G, \delta) $.
Hence,
$
m_{\rm SE}(v_E,s_E,\delta)
\le
m_{\rm SG}(v_G,\delta)
$
holds if and only if
$
\sqrt{2v_E L}
\le
\sqrt{2v_E\gamma L}
$,
which is equivalent to
$
\gamma\ge1
$.

Next, consider the case
$
L>\frac{d^2}{2}
$.
Since
$
s_E=\frac{\sqrt{v_E}}{d}
$, $ m_{\rm SE} $ can be written as
\begin{align*}
m_{\rm SE}(v_E,s_E,\delta)
&=
s_E L+\frac{v_E}{2s_E}=
\sqrt{v_E}
\left(
\frac{L}{d}+\frac{d}{2}
\right).
\end{align*}
Therefore, by \eqref{eq:m_SG1},
$
m_{\rm SE}(v_E,s_E,\delta)
\le
m_{\rm SG}(v_G,\delta)
$
is equivalent to
\[
\frac{L}{d}+\frac{d}{2}
\le
\sqrt{2\gamma L}.
\]
Since both sides are nonnegative, squaring both sides gives
\[
\left(
\frac{L}{d}+\frac{d}{2}
\right)^2
\le
2\gamma L,
\]
or equivalently,
\[
\gamma
\ge
\frac{
\left(
L/d+d/2
\right)^2
}{
2L
}.
\]
This completes the proof.
\end{proof}

Proposition~\ref{prop:dominance} shows that the relative tightness of the two bounds depends on the variance proxies and the prescribed probability level. When \(v_E<v_G\), the sub-exponential bound can be tighter for less stringent confidence levels. For sufficiently small \(\delta\), however, the sub-Gaussian bound becomes tighter, since its margin grows as \(\sqrt{\log(1/\delta)}\), compared with \(\log(1/\delta)\) for the sub-exponential bound.
Thus, the proposition provides a criterion for selecting the less conservative bound for a given constraint direction and probability level.

\section{Numerical illustration}
\label{sec:numerical}

We illustrate the comparison between the sub-Gaussian (SG) and
sub-exponential (SE) propagation schemes using an example adapted from \cite{Moussa2026}
\begin{equation}
A(\xi_k)=\begin{bmatrix}
\xi_{k,1}&0&\xi_{k,1}\\
0.1&0.1+\xi_{k,2}^2-\sigma_2^2&0.2\\
0.2&0.1&0.5+\xi_{k,2}
\end{bmatrix},
\label{eq:numerical_A}
\end{equation}
where $\sigma_2^2:=\mathbb E[\xi_{k,2}^2]$. The components of $\xi_k$ are given by
\[
\xi_{k,1}=0.08Z_{k,1},\qquad
\xi_{k,2}=0.21Z_{k,2},
\]
where $Z_{k,i}$ are independent and follow the standard Gaussian distribution truncated to $[-2,2]$.
Thus, $\Xi_k=[-0.16,0.16]\times[-0.42,0.42]$.
We take $x_0\sim\mathcal N(\mu_0,10^{-3}I_3)$, $\mu_0=[0.35\;0.25\;0.10]^\top$, and $w_k\sim\mathcal N(0,10^{-2}I_3)$, where $\calN(\mu,\Sigma)$ denotes the Gaussian distribution with mean $\mu$ and covariance matrix $\Sigma$. The initial state $x_0$, all $w_k$, and all $\xi_k$ are mutually independent.

Let $e_i$ denote the $i$th standard basis vector, and define $G_1:=e_1(e_1+e_3)^\top$, $G_2:=e_2e_2^\top$, and $G_3:=e_3e_3^\top$.
Then,
\begin{equation*}
A(\xi_k)=\bar A_k+\xi_{k,1}G_1
+(\xi_{k,2}^2-\sigma_2^2)G_2+\xi_{k,2}G_3.
\end{equation*}
By combining this with Hoeffding's lemma, it can be shown that
$a_k\equiv a:=\max\! \left\{
 \frac{0.16^2}{\sigma_1^2},\,
 \frac{M_q^2}{\sigma_q^2}+\frac{0.42^2}{\sigma_2^2}
 \right\}$ satisfies Assumption~\ref{ass:A_subgauss}, where $\sigma_1^2 := \bbE[\xi_{k,1}^2]$, $\sigma_q^2 := \bbE[\xi_{k,2}^4] - (\bbE[\xi_{k,2}^2])^2$, and $M_q:=\max\{\sigma_2^2,\,0.42^2-\sigma_2^2\}$.
The coefficients multiplying $G_1,G_2,G_3$ lie in
$[-0.16,0.16]$, $[-\sigma_2^2,0.42^2-\sigma_2^2]$,
and $[-0.42,0.42]$, respectively.
Varying these coefficients independently
over their respective intervals yields eight vertex
matrices $A^{(i)}$, $i=1,\ldots,8$, such that
$
A(\xi)\in\operatorname{conv}
\{A^{(1)},\ldots,A^{(8)}\}
$ for any $\xi\in\Xi_k$.
Then, we have
\begin{equation}
    \beta := \max_{i=1,\ldots,8}\|A^{(i)}\|_2^2
    \simeq 0.99<1 .
    \label{eq:beta}
\end{equation}

Although $A(\xi)$ is not affine in $\xi$, the above
convex-hull representation allows us to apply
the vertex-based argument in Remark~\ref{rmk:vertices}.
Thus, for the sub-Gaussian propagation, we apply Proposition~\ref{prop:subg_propagation} by solving the following semidefinite program (SDP):
\begin{align}
\min_{\Sigma_{G,k+1}}\quad &\operatorname{trace}(\Sigma_{G,k+1}) \label{eq:numerical_sg}\\
\mathrm{s.t.}\quad
&\Sigma_{G,k+1}\succeq A^{(i)}\Sigma_{G,k}A^{(i)\top}
+aH_k+10^{-2}I_3, \nonumber\\
&i=1,\ldots,8, \nonumber
\end{align}
where $H_k:=\operatorname{cov}(A(\xi_k)\mu_k)$ and
$\mu_{k+1}=\bar A_k\mu_k$.
The SDP~\eqref{eq:numerical_sg} is solved sequentially, starting from
$\Sigma_{G,0}=10^{-3}I_3$.
As a simpler alternative based on Proposition~\ref{prop:subg_propagation}, we also consider an isotropic sub-Gaussian variance proxy of the form $\Sigma_{k}=g_k I_3$, propagated according to
\begin{equation}\label{eq:isotropic_propagation}
\begin{aligned}
g_{k+1}
&=\beta g_k+a\lambda_{\max}(H_k)+10^{-2},
\end{aligned}
\end{equation}
where $g_0=10^{-3}$ and $\beta$ is given in~\eqref{eq:beta}.
Since $A^{(i)}(g_k I_3)(A^{(i)})^\top \preceq\beta g_k I_3$ for all vertices, this recursion satisfies Proposition~\ref{prop:subg_propagation}.

For the sub-exponential propagation, we use Theorem~\ref{prop:subexp_propagation} with
$\Sigma_{E,0}=10^{-3}I_3$ and choose $S_{k+1}$
by Proposition~\ref{prop:robust_S_design}, starting from $S_0=0.038I_3$
and using $\nu=18$. 
For comparison, we also propagate the exact covariance $C_k=\operatorname{cov}(x_k)$ and apply Cantelli's inequality. The resulting margin $m_{\mathrm{C}}(k,\delta) =\sqrt{(1-\delta)h^\top C_kh/\delta}$ satisfies $\mathbb{P}\{ h^\top(x_k-\mu_k)>m_{\mathrm{C}}(k,\delta) \}\leq\delta$ for any $\delta \in (0,1)$.

\begin{figure}[t]
    \centering
    \begin{subfigure}[t]{0.49\linewidth}
        \centering
        \includegraphics[width=\linewidth]{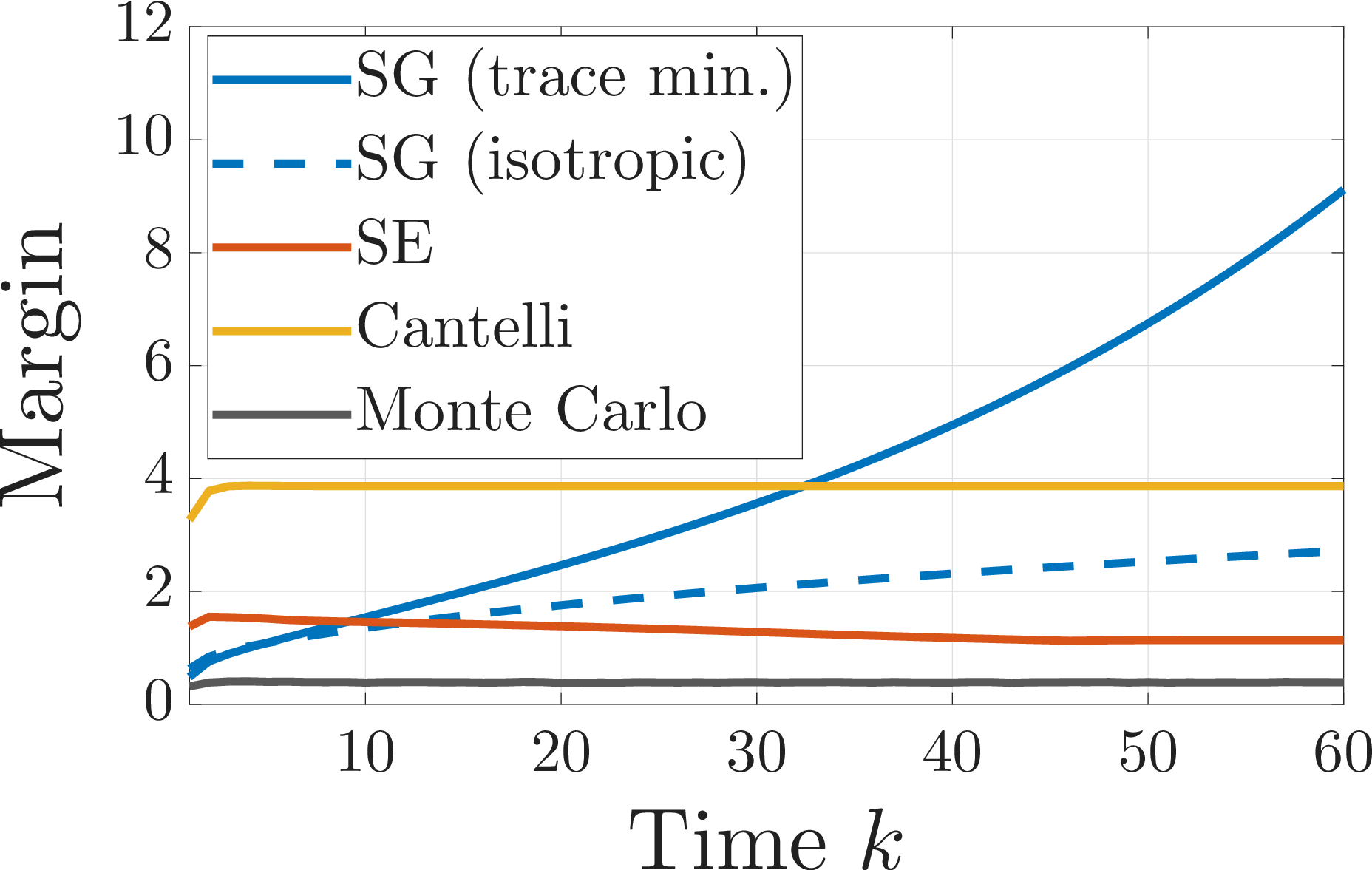}
        \caption{Multi-step propagation}
        \label{fig:numerical_multistep}
    \end{subfigure}
    \hfill
    \begin{subfigure}[t]{0.49\linewidth}
        \centering
        \includegraphics[width=\linewidth]{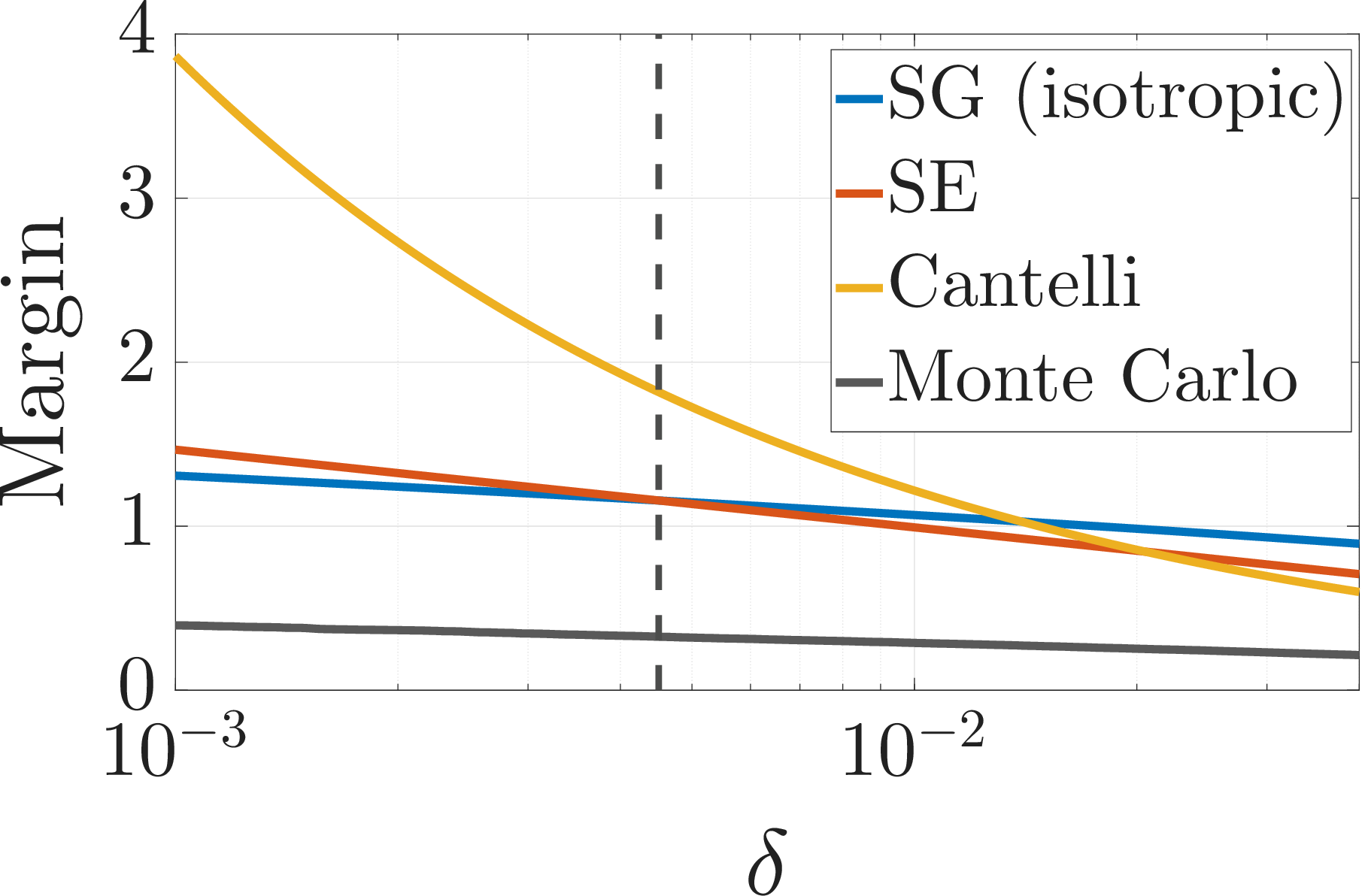}
        \caption{Dependence on $\delta$}
        \label{fig:numerical_delta}
    \end{subfigure}

    \caption{Comparison of half-space bounds for $h = e_3$.
    (a) Multi-step propagation with $\delta=0.001$, showing SG with sequential trace minimization~\eqref{eq:numerical_sg} (solid blue) and isotropic propagation~\eqref{eq:isotropic_propagation} (dashed blue).
    (b) SG (isotropic), SE, and Cantelli margins, and Monte Carlo empirical quantiles
as functions of $\delta$ at $k=9$;
the vertical line indicates the intersection
of the SG and SE margins.}
    \label{fig:numerical_comparison}
\end{figure}

Fig.~\ref{fig:numerical_multistep} compares the
half-space margins for $h=e_3$ and $\delta=0.001$ over $60$ steps, together with empirical quantiles from $10^5$ Monte Carlo trajectories.
The SG margin obtained by the sequential trace minimization~\eqref{eq:numerical_sg} increases substantially over the horizon, whereas the isotropic SG margin grows much more slowly and is guaranteed to converge because $\beta<1$ and $H_k\to0$.
The SE margin approaches a constant.
These results illustrate the advantage of expectation-based SE propagation in this example, while also highlighting the sensitivity of SG propagation to the choice of variance proxy.

Fig.~\ref{fig:numerical_delta} compares the analytical margins with Monte Carlo empirical quantiles for different violation probabilities.
The SG and SE margins at $k=9$ intersect near $\delta=0.0045$, consistently with the dominance condition in Proposition~\ref{prop:dominance}. The SG bound is tighter for
smaller $\delta$, whereas the SE bound is
tighter for larger $\delta$.
This result highlights the complementary nature of the SG and SE characterizations.
Cantelli's bound becomes particularly conservative
for small $\delta$.

The Monte Carlo empirical quantiles in Figs.~\ref{fig:numerical_multistep} and \ref{fig:numerical_delta} do not provide certified probability guarantees.
Moreover, their computation relies on the fully specified probability distributions of $x_0$, $\xi_k$, and $w_k$ to generate sample trajectories. 
In contrast, the SG and SE propagation schemes do not require full knowledge of these distributions, relying instead on suitable support, concentration, and moment information.

\section{Conclusion}\label{sec:conclusion}
This paper addressed probabilistic reachability for linear
stochastic systems subject to multiplicative uncertainty and
additive noise. Two complementary propagation approaches
were developed based on sub-Gaussian and sub-exponential
characterizations, allowing the construction of ellipsoidal
and half-space probabilistic bounds. 
Numerical results illustrated their complementary behavior
depending on the prescribed probability level.
Future work will focus on the incorporation of these bounds into SMPC, including recursive feasibility under stochastic parametric uncertainty. The framework of Knaup and
Tsiotras~\cite{Knaup2026} for time-varying systems with unbounded
additive noise provides a possible starting point.

\appendix

\textit{Proof of Proposition~\ref{prop:A_subgauss_affine}:}
For any \(v\in\bbR^n\), define
$
    B_v
    :=
    \begin{bmatrix}
        \mathcal{A}_1v & \cdots & \mathcal{A}_pv
    \end{bmatrix}
    \in\bbR^{n\times p}
$.
Since \(\bbE[\xi_k]=0\), one has
\(
    \bar A_k
    =
    \bbE[A(\xi_k)]
    =
    \mathcal{A}_0,
\)
and therefore
\(
    \left(
        A(\xi_k)-\bar A_k
    \right)v
    =
    B_v\xi_k.
\)
Using the sub-Gaussian characterization of \(\xi_k\), for any \(\lambda\in\bbR^n\),
\[
    \bbE
    \! \left[
        \exp\!\left(
            \lambda^\top B_v\xi_k
        \right)
    \right]
    \leq
    \exp\!\left(
        \frac{1}{2}
        \lambda^\top
        B_v\Sigma_{\xi,k}B_v^\top
        \lambda
    \right).
\]
Moreover,
\[
    {\rm cov}(A(\xi_k)v)
    =
    \bbE
    \! \left[
        B_v\xi_k\xi_k^\top B_v^\top
    \right]
    =
    B_v\Gamma_{\xi,k}B_v^\top.
\]
By definition of
\(
    a_k
    =
    \lambda_{\max}
    \! \left(
        \Gamma_{\xi,k}^{-1/2}\Sigma_{\xi,k}\Gamma_{\xi,k}^{-1/2}
    \right),
\)
we have
\(
    \Sigma_{\xi,k}
    \preceq
    a_k\Gamma_{\xi,k}.
\)
Consequently,
\[
    B_v\Sigma_{\xi,k}B_v^\top
    \preceq
    a_kB_v\Gamma_{\xi,k}B_v^\top
    =
    a_k\,{\rm cov}(A(\xi_k)v).
\]
Hence,
\[
    \bbE
    \! \left[
        \exp\!\left(
            \lambda^\top
            \left(
                A(\xi_k)-\bar A_k
            \right)v
        \right)
    \right]
    \leq
    \exp\!\left(
        \frac{1}{2}
        \|\lambda\|^2_{
            a_k\,{\rm cov}(A(\xi_k)v)
        }
    \right),
\]
which completes the proof.

\section*{Acknowledgment}
The authors acknowledge the use of ChatGPT (OpenAI) for language editing and text reformulation. All content was carefully reviewed and verified by the authors, who take full responsibility for the final manuscript.

\bibliographystyle{unsrt}   
\bibliography{Biblio} 
\end{document}